# Development of Virtual Accelerator Environment for Beam Diagnostics[*]

Gu Duan, Zhang Meng, Gu Qiang, Huang Dazhang, Zhao Minghua

(*Shanghai Institute of Applied Physics, Chinese Academy of Sciences, Shanghai 201800, China*;)

**Abstract:** For the proper operation of Dalian Compact Light Source (DCLS) linac, measurement and control of the electron bunch is critical. In order to test control algorithms and high level physical applications, a virtual accelerator environment is constructed based on ELEGANT and SDDS toolkit. The required beam parameter measurement and orbit control tools are verified under this virtual environment and it will greatly speed up the development of machine commissioning tools. The design concept and current development status are presented.

Dalian Coherent Light Source is a free electron laser user facility based on the principle of single-pass, high-gain harmonic generation scheme[1]. During the commissioning of the linac section, beam parameter measurement is important, meanwhile, beam correction is indispensable for the stable operation of the machine. To expedite the design of measurement tools, Virtual Environment (VA) has been widely used on many facilities[2-6]. Usually, beam energy and energy spread are measured by a diagnostic system that consists of a bending magnet and the following beam profile monitor, and the transverse beam emittance is measured using the so-called "Quadrupole Scan" method. Beam orbit is measured by Beam Position Monitor (BPM) along the beamline and the orbit correction is implemented using the dipole correctors. Resolution and accuracy of the beam instrument should be specified during the design stage and analysed during the commissioning stage.

In order to evaluate the feasibility of beam diagnostic systems, a virtual accelerator environment based on ELEGANT[7] and SDDS toolkit is developed. With the tools and interfaces provided by this VA, the beam behaviours simulated by physics codes can be directly handled to High Level Applications (HLA), this will greatly speed up the development of machine commissioning tools.

In its first part, this paper focuses on the concept of the virtual accelerator environment and its role during the communication between physics codes and high level applications. In the second part, with the flexibility of this environment, the energy and energy spread measurement tools based on it are shown, which can be used to find the proper RF phases before and after chicane. In the third part, beam emittance measurement tools based on VA is shown. Using this tool, the transverse parameters which give the information for the beamline can be derived. The fourth part describes the design of the beam orbit correction tool and the response matrices acquirement both theoretically and experimentally. In the final section a short conclusion is given.

[*] Received date: ;　　　　　Revised date:
Foundation item: Supported by Knowledge Innovation Program of CAS and Natural Science Foundation of Shanghai City, Grant No. 12ZR1436600.
Biography: Gu Duan (1982—), male, PhD candidate, engaged in accelerator physics and diagnostics; e-mail: guduan@sinap.ac.cn
Corresponding author: Zhao Minghua (1961—), male,; e-mail: zhaominghua@sinap.ac.cn.

## 1  Virtual accelerator environment based on ELEGANT and SDDS toolkit

SDDS toolkit and ELEGANT code have been extensively used during the design of Dalian Compact Light Source. Integrating physics codes into control system will give great potential to optimize the machine commissioning and operation. So we have developed a Virtual Accelerator to provide tools for integrating and interfacing different simulation codes into the control system. The VA uses ELEGANT code as the simulator of beam dynamics to simulate the behaviour of electron beam, meanwhile, with the combination of SDDS toolkit, the VA can also provide data pre-processing and exchanging for different simulation codes. As the control system is based on Experiment Physics and Industrial Control System (EPICS)[8], the communication and data transfer are provided by using the EPICS portable Channel Access (CA), so a set of utilities for interfacing the VA and the high level control system have been developed. In this way, the VA will operate just like the real machine from the view of EPICS channel. The structure of the Virtual Accelerator compared with real machine is shown in Fig. 1.

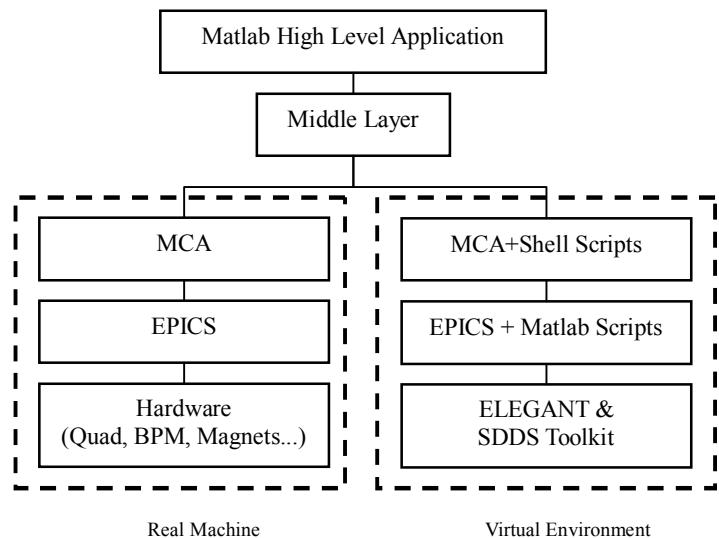

Fig.1  The structure of the virtual accelerator compared with real machine

In the case of DCLS, the High Level Applications (HLA) including Graphical User Interface (GUI) has been designed on MATLAB platform. In order to enable the communication between HLA and the physics simulation codes, an I/O interface is developed which can convert files between the SDDS format and the Matlab format. With the flexibility of the built-in functions in SDDS toolkit, the SDDS format file can be loaded directly into Matlab HLA. An example of reading and display the output beam distribution from simulation is shown as Fig. 2.

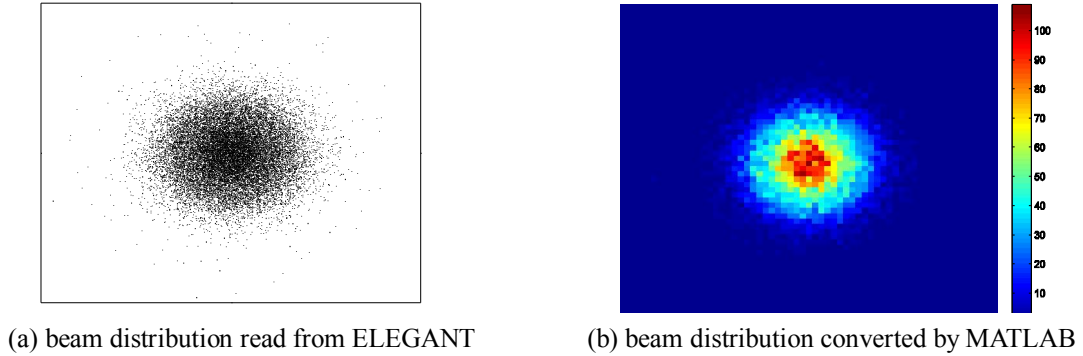

(a) beam distribution read from ELEGANT    (b) beam distribution converted by MATLAB

Fig. 2.  Simulated and converted beam distributions

For the reverse path, the parameters and configurations from HLA can be passed to physics simulation codes via the specified function. The conversion between machine quantities (e.g. current in Ampere) and physics quantities in simulation codes (e.g. quadrupole strength K in m$^{-2}$) are calculated and stored in advance so the VA can be operated like an online machine.

## 4    Beam parameter measurement in longitudinal direction

With the flexibility of VA environment, we can measure and tune the longitudinal beam parameters for the DCLS linac. According to the design layout, beam energy and energy spread measurement are routinely carried out at three beam diagnostics locations: at the exit of RF gun, at the exit of ACC2 and at the linac end. The basic layout of bending magnet system is shown in Fig. 3.

The beam energy can be evaluated from the magnetic field of the bending magnet as described in Eq. 1.

$$E[MeV] = 299.8 \cdot B\rho[T \cdot m] \quad (1)$$

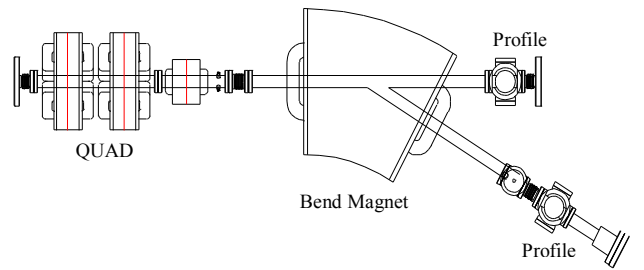

Fig. 3. The basic layout of bending magnet system

where $\rho$ is the bending radius which is fixed by the geometric distance between profile and bending magnet, $B$ is the magnetic field of the bending magnet which is controlled by current[9]. In order to simulate the usual procedure of energy measurement, during the calculation of the beam distributions using simulation code, each location equipped with one "watch point" element, by which we can get the theoretic 3D beam distribution. As the beam energy distribution is characterized by analyzing the transverse profile collected on a profile, only the x-y plane beam information are handled to reconstruct the beam transverse distribution. The beam shape and center are fitted using Gaussian fit method and shown on the HLA panel. As described in Eq.1, the beam energy can be determined as long as the current of bending magnet is measured.

Meanwhile the energy spread measurement is carried out using the same structure. The energy spread is derived by measuring the beam size at the profile screen and acknowledging the local dispersion[10]. The transverse beam size $\sigma_{beam}=\sqrt{\sigma_\beta^2+\sigma_\delta^2}$ is measured in a dispersive location has two sources. where $\sigma_\beta=\sqrt{\varepsilon\beta}$ is the beam's betatron size and $\sigma_\delta$ is the size due to dispersion. So the energy spread is $\sigma_E=\sigma_\delta \cdot D/E_{beam}$. Where $D$ is the dispersion. when $\sigma_\beta \ll \sigma_\delta$, the betatron contribution can be ignored, then the energy spread can be dominated by transverse beam size and dispersion only. By using the strategy above, the upstream quadrupoles are optimized to get the minimum betatron contribution after 30 degree bending magnet, then the according energy spread is determined[11]. The final energy spread measurement interface is shown as Fig 4.

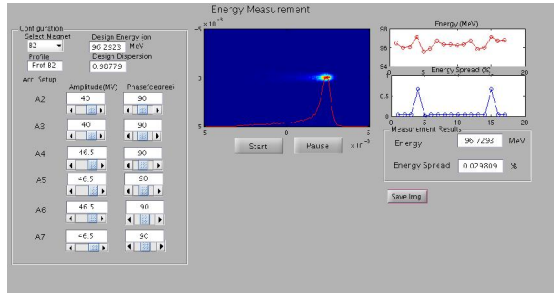

Fig. 4. The final energy spread measurement interface

The linac is operated off-crest to chirp the bunch for longitudinal compression. In practice, the accelerate phase is indicated by the offset of minimum energy spread from maximum energy[12]. Using energy and energy spread data from previous measurements, the required RF phase of subsequent accelerate modules can be found where the energy spread is the smallest as shown in Fig. 5.

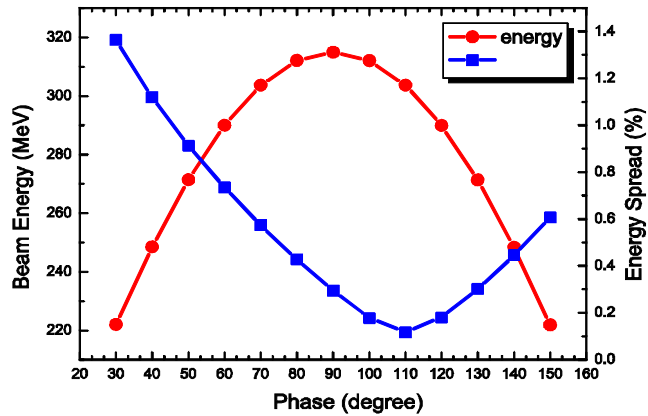

Fig. 5. Seek the required RF phase of subsequent accelerate modules.

## 3  Beam emittance measurement in transverse direction

As one of the most important judgments of beam quality, projected emittance measurement is critical for optics matching in linac. Many different ways have been studied to measure the emittance, in the case of DCLS, an automatic measurement tool is developed based on VA using "Quadrupole Scan" method[13]. During the scanning, with assuming the thin lens approximation, the correlation between $\sigma_s^2$ and $K \cdot L_q$ is given as Eq. 2.

$$\sigma_s^2 = A(K \cdot L_q)^2 - 2A \cdot B(K \cdot L_q) + (A \cdot B^2 + C) \tag{2}$$

Where $\sigma_s$ is the rms beam size on the profile screen, $L_q$ is quadrupole length, $K$ is the normalized quadrupole strength, $A$, $B$ and $C$ are coefficients that can be fitted by the second order polynomial fitting. The normalized projected emittance $\varepsilon_n$ can be written as $\varepsilon_n = \sqrt{AC}/L_d$. Where $L_d$ is the length of drift between quadrupole and profile screen. There are five reserved diagnostics locations for emittance measurement and each location equipped with one profile, after scanning the upstream quadrupole, the emittance measurement tool can extract the transverse beam size on profile and estimate the transverse beam emittance at the entrance of the scanned quadrupole. The emittance will be determined for both horizontal and vertical plane simultaneously. At the same time, twiss parameters at the entrance of scanned quadrupole can also be measured. Space charge forces are not included in the emittance analysis and should have no significant impact at these energies[14]. During the tracking in ELEGANT, 100000 particles are used, when the beam image converted from beam distribution loaded, it is shown in a $10mm \times 10mm$ area, the pixel calibration is $10um/pixel$. In order to remove the influence of background noise, an averaged background image is subtracted from each beam image in Matlab procession. When the emittance measurements are performed sequentially at each diagnostic location, useful information about the emittance transport through the linac can be obtained[15].

An example of emittance measurement software interface and result is shown as Fig. 6. In which the scanned quadrupole values versus observed beam size square are measured and then fitted. Finally the emittance of $1mm \cdot mrad$ were determined from the fitted curves for both horizontal and vertical planes.

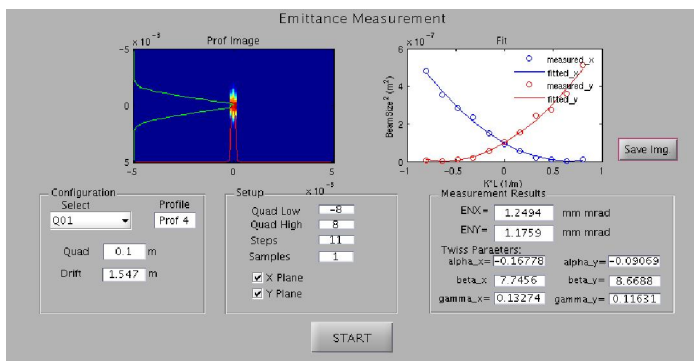    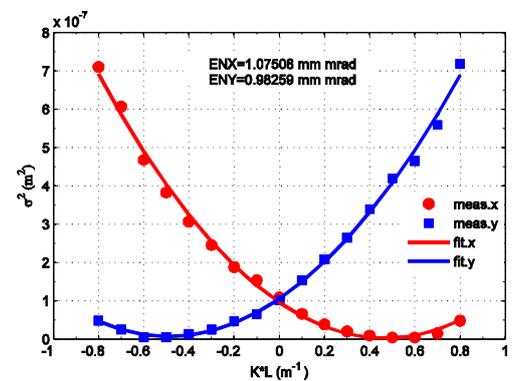

(a) emittance measurement interface  (b) emittance measurement example result

Fig. 6. Emittance measurement software interface and result example.

## 4  Beam orbit correction

With the existence of element misalignments, the orbit need to be corrected. Usually, a common used SVD based orbit algorithm like Global method is applied on the linac. Beam orbit correction is accomplished using BPM readings and in turn correct the orbit with dipole magnets. Correction tools has been developed and debugged under the VA environment. This algorithm relies mainly on a so-called Response Matrices which maps the orbit changes due to the corrector kicks. The main issue when applying orbit correction is how to get the according response matrices for both planes. Fortunately, the theoretical response matrices can be computed by ELEGANT and imported into a Matlab Orbit Correction tool. Experimentally, the response matrices can be determined by measuring the BPM readings while varying the current of each dipole corrector. Compared with the time consuming experimental measurement, using the theoretical response matrices is much faster and can be applied to the orbit control system without affecting the facility operation. Considering the demand of verify the consistency between the theoretical and experimental model.

Fig. 7 shows the typical improvement in beam position at each BPM achieved by the correction tool. In this example, each quadrupole is assumed with $100\mu m$ transverse offset with respect to a straight line. In the horizontal plane the RMS error is reduced from $0.41mm$ to $0.06mm$ while in the vertical plane the RMS error is reduced from $0.52mm$ to $0.05mm$.

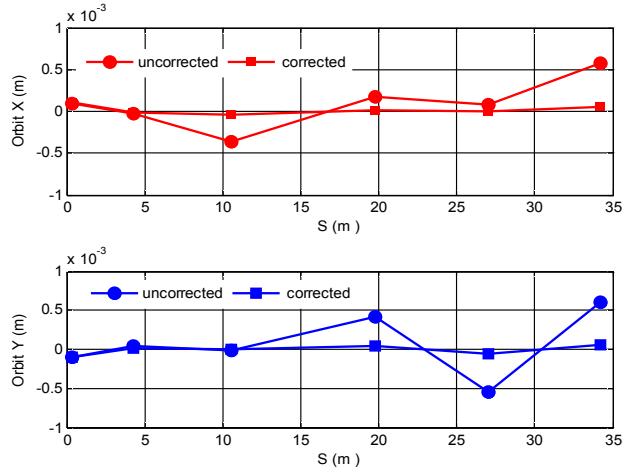

Fig. 7. Improvement after orbit correction for both vertical and horizontal planes.

## 5  Conclusion

A set of beam parameter measurement and orbit control tools based on virtual accelerator environment have been developed. The VA environment provides flexible data process interfaces with physics simulation codes. The theoretical model from simulation can be directly imported into high level applications or through EPICS channel access. Different beam measurement procedures and correction methods can be tested and verified benefited from this VA environment. Successful integrating this VA environment into the control system will greatly improve the machine performance in the DCLS commissioning.

# .束流诊断虚拟加速器环境开发

谷端， 张猛， 顾强，黄大章，赵明华

（中国科学院上海应用物理研究所，上海，201800）

**摘 要**： 在大连紧凑型自由电子激光装置直线加速器的调试和运行中，束流能量、发射度和轨道控制是至关重要的。为了更好的测试控制方法和开发高层控制软件，构建了一个基于 ELEGANT 和 SDDS Toolkit 的虚拟加速器环境。本文描述了束流参数测量和轨道控制等软件在该虚拟加速器环境中的开发和测试，并探讨了该虚拟加速器的设计理念和未来可能的应用。

**关键词**：虚拟加速器环境；束流参数测量；上层控制软件；